# Virtual photons in macroscopic tunnelling

H. Aichmann [1], G. Nimtz [2*], P. Bruney[3]

**Affiliations:**
[1] Zur Bitz 1, 61231 Bad Nauheim, Germany.
[2] II. Physikalisches Institut, Universität zu Köln, Zülpicherstrasse 77, 50937 Köln, Germany.
[3] Plane Concepts Inc., Silver Spring, Maryland 20904 USA.

Abstract − Tunnelling processes are thought to proceed via virtual waves due to observed superluminal (faster than light) signal speeds. Some assume such speeds must violate causality. These assumptions contradict, for instance, superluminally tunnelled music and optical tunnelling couplers applied in fiber communication. Recently tunnelling barriers were conjectured to be cavities, wherein the tunnelled output signal is not causally related with the input. The tests described here resolve that tunnelling waves are virtual, propagations are superluminal, and causality is preserved.

Tunnelling and its optical equivalent, the evanescent mode, have been conjectured to involve virtual particles and waves [1-4]. Evanescent modes are described by imaginary wavenumbers. They are not measurable inside a barrier. In order to make them evident, either the energy of the virtual particle or wave has to be increased up to the barrier's height or the barrier has to be reduced [5]. Of course, during either procedure, the virtual character of the particles or waves is lost. Previous experimental tunnelling studies with photons and phonons indicated that the virtual waves or particles are non-local and thus spread instantaneously; the barrier traversal time is zero [4]. Zero time within the barrier and the interaction time at the barrier front yield an overall superluminal velocity (see, e.g., Refs. [10, 6, 7]). However, a superluminal physical signal velocity does *not* necessarily violate causality [8, 9]; effect does *not* precede

(a) E-mail: G.Nimtz@uni-koeln.de

cause. In this study we investigated the propagation of virtual photons by inductive posts in undersized waveguides and in dielectric quarter wavelength lattices. The impedances of barriers were obtained by measuring the Voltage Standing Wave Ratio.

Evanescent modes are expected in optics in the case of total reflection, or in internal frustrated total reflection where, for example, two double prisms or two fiber cables in optical communications are coupled. They are field solutions of the Helmholtz equation, which is mathematically identical with the Schrödinger equation. The tunnelling solutions of the Schrödinger equation also have imaginary wave numbers. Sommerfeld pointed out that the optical evanescent modes represent an analog of the quantum mechanical tunnelling modes [11]. He used double prisms as a tunnelling example (displayed in Fig. 1a). After several earlier calculations on the barrier traversal time of particles, Hartman [10] calculated tunnelling of wave packets. He was motivated by the discovery of the (tunnelling) Esaki diode and by tunnelling across thin insulating layers at that time. Hartman deduced that, for opaque barriers, a short interaction time is needed at the barrier front. Remarkably, this traversal time was constant and independent of barrier length for opaque barriers. This strange effect, named after Hartman, was actually observed in several analog experimental studies [12-14].

Experiments:

The optical barriers tested here are shown in Fig.1.

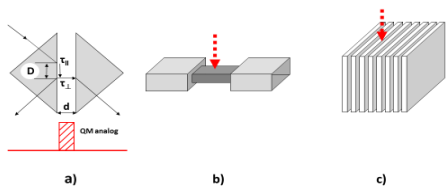

**Fig. 1:** Three barrier types: a) Double prisms. Distances D and d equal the Goos-Hänchen shift and the potential barrier length respectively. $\tau_{||}$ and $\tau_{\perp}$ are the propagation times parallel and normal to the prisms, respectively. b) Undersized waveguide with a barrier length of 40 mm. The frequency cut-off at a low frequency of 9.49 GHz is given by the narrow Ku-band waveguide width (15.8 mm), whereas the larger waveguide X-band has a cut-off frequency of 6.49 GHz (width 22.86 mm). c) Quarter wave length dielectric lattice. Lattice of 8 quarter wavelength Perspex layers (n = 1.6) was investigated. Red arrows indicate the locations of the inserted inductive post.

According to quantum mechanics, a particle with mass $m$ observed in the exponential tail of the tunnelling probability must be localized within a distance in the order of $\Delta z \approx 1/\kappa = 1/ik$, where k and $\kappa$ are the real and imaginary wave numbers, respectively. Hence, its momentum must be uncertain by $\Delta p > \hbar/\Delta z \approx \hbar\kappa = [2m\, (U_0 - W_{kin})]^{1/2}$, where $U_0$ is the barrier potential. The particle of energy $W_{kin}$ can thus be located in the non-classical region only if it is given an energy $U_0 - W_{kin}$ sufficient to raise it into the classically allowed region [5]. In analogy this holds for electromagnetic waves as is shown in this study. For instance, in a slotted waveguide, a standing wave can be probed through a tiny slot as long as the frequency is higher than the cut-off frequency of the waveguide. Evanescent modes with their imaginary wave numbers are not expected to be detectable below the cut-off frequency of the waveguide. In general, the dielectric barrier in optics has to be matched to the dielectric medium of the real photon state. There, the potential $U_0$ corresponds to $n^2$, where *n* is the refractive index of the dielectric medium. In this study the match was achieved by a so-called inductive post which was used as field probe as well [15].

The resonator of a LASER is usually designed with metallic or dielectric mirrors. Metallic mirrors are lossy, whereas mirrors based on evanescent (i.e. tunneling) modes are elastic barriers. The latter property is the reason for their application in high power LASER systems. In special cases, e.g. at microwave frequencies, high pass filters or so called frequency cut-off waveguides may be applied as mirrors. The mirror barriers are sketched in Fig. 1b, c. The smaller reflectivity and the higher losses of metallic mirrors at wavelengths below 1 mm present a handicap in the case of high power LASERs [16].

In this study we compare the complex reflectivity and the transmission behaviour of two mirrors at microwave frequencies. The reflectivity of metals is caused by free carriers, whereas that of dielectric mirrors is based on Bragg reflection. In high pass filters, waves have an imaginary wave number below their cut-off frequency like dielectric mirrors have in their forbidden band gap.

Dielectric mirrors are usually 1-dimensional lattice structures as shown in Fig. 1c, where the lattice is built by quarter wavelength layers that differ periodically in their refractive index. A frequency cut-off waveguide is sketched in Fig.1b. Waves with frequencies below the cut-off value of the narrow guide become evanescent; i.e. the wave number is imaginary and thus the wave propagation vanishes and there is no real wave propagation. The cut-off frequency (i.e. the dielectric barrier height) is determined by the geometry of the guide. For example, special waves ($H_{10}$-mode) become evanescent if their wavelength exceeds twice the waveguide width.

The phase shift of the reflected beam at the mirror surface was calculated from the measured geometric shift of the Voltage Standing Wave Ratio (VSWR). The shift of the reflected EM waves is $\pi$ at a surface with a higher refractive index or at a metal surface. The $\pi$ step also takes place for quantum mechanical wave functions at a potential barrier.

Quarter wavelength lattices, high pass mirrors, and double prisms exhibit a universal reflection time and virtual transmission in electromagnetic and elastic fields as demonstrated in Table 1.

The experimental set up is shown in Fig. 2. The standing wave patterns of the reflection were measured at microwave frequencies near 10 GHz, i.e. at wavelengths of about 30 mm (this frequency range is called the microwave X-band). Barriers of the mechanical smaller Ku-band waveguide were also studied; they have a cut-off frequency of 9.49 GHz (the corresponding wavelength is 31.6 mm). We measured the VSWR with a slotted X-band line above the cut-off frequency of Ku-band. In addition, the absolute value of the reflectivity R, was measured with a directional coupler power meter combination for comparison.

The dielectric potential shift of both the undersized waveguide and the dielectric lattice were obtained by using an inductive post [14]. The post was inserted into the waveguide as shown in Figs. 1b and 1c. It acts as an inductivity depending on its geometrical dimensions and position inside the waveguide. We used the core of a small semi rigid coax cable with a diameter of 0.4 mm. This cable was connected to a detector. With a small penetration into the waveguide (less than 2 mm), the post did not influence the transmission above noise and did not act as an antenna. However, at deeper penetrations into the waveguide or the lattice (in its third air layer), the waveguide impedance matches the regular waveguide impedance. In this state, the post becomes excited, thus the transmission is increased and the reflection decreased. Examples are displayed in Figs. 3, 4.

| Tunnelling barriers | Reference | $\tau$ | T = 1/$\nu$ |
|---|---|---|---|
| *Frustrated total reflection* | [4] | 117 *ps* | 120 *ps* |
| *Double prisms* | [19] | 87 *ps* | 100 *ps* |
| *Photonic lattice* | [20] | 2.13 *fs* | 2.34 *fs* |
| *Photonic lattice* | [21] | 2.7 *fs* | 2.7 *fs* |
| *Undersized waveguide* | [22] | 130 *ps* | 115 *ps* |
| *Electron field-emission tunnelling* | [23] | 7 *fs* | 6 *fs* |
| *Electron ionization tunnelling* | [24] | < 1 *as* | ? *as* |
| *Acoustic (phonon) tunnelling* | [25] | 0.8 *μs* | 1 *μs* |
| *Acoustic (phonon) tunnelling* | [25] | 0.9 *ms* | 1 *ms* |

**Table 1:** For comparison: Reflection time (equals transmission time for symmetric barriers) at various barriers (mirrors) and electric and elastic fields. The studied frequency range is 1 kHz to 427 THz. $\tau$ is the measured barrier traversal time, and T the empirical universal time, T = 1/$\nu$, which equals the oscillation time of the tunneling wave [4, 18]. A theoretical analysis and fundamental base of the universal tunnelling time is presented in Ref. [17].

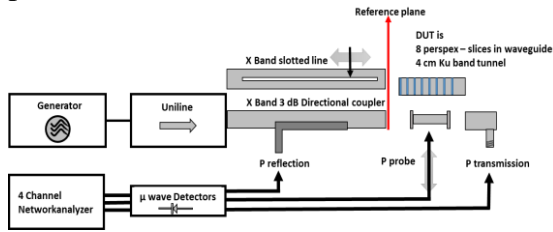

**Fig. 2:** The set up to measure transmission, reflection, probe power, VSWR and phase with an X-band waveguide slotted line, and alternatively to measure return loss with a directional coupler. Measurement positions: in front, inside, and behind the barrier. Frequency: 8.0 – 10.5 GHz; probe 0.4 mm $\phi$; probe length 8 mm; hole diameter 1 mm. **D**evices **U**nder **T**est are the different barriers.

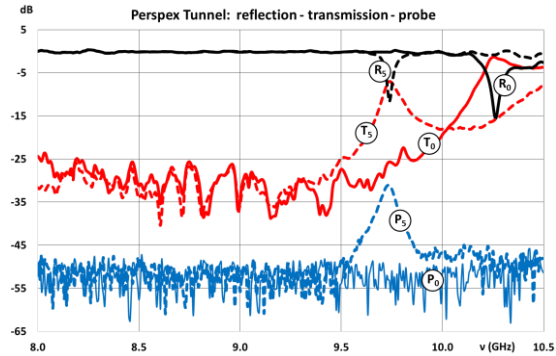

**Fig. 3:** Measurement results for transmission T, return loss R and Probe signal P (at 0 and 5mm probe depths) for the quarter wavelength Perspex/air tunnel.

The dielectric mirror of periodical quarter wavelength layers with alternating real refractive index displays a VSWR of 200 ± 20 with an inductive phase shift of + 6.5°. The studied mirror was an 8 Perspex/air quarter wavelength lattice (Fig.1c). The 8 layers were built within an

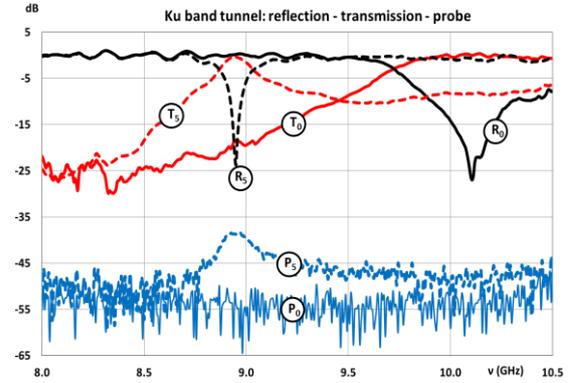

**Fig. 4:** Measurement results for transmission T, return loss R and probe signals P (at 0 and 5mm probe depths) for the Ku-band tunnel (tunnel length = 40 mm).

X-band waveguide. Layer thicknesses 6.1 mm Perspex (n = 1.6 and 12.0 mm air). The corresponding interaction time of ≈ $1/\nu$ in time domain and frequency domain studies (Table 1) did not influence the VSWR, where $\nu$ is the frequency. The undersized waveguide high pass mirror (Fig. 1b) also shows a VSWR of 200 ± 20, but displays a capacitive phase shift of -38°.

In Figs. 3, 4 transmission T, reflection R, and probe signals P are shown for the high pass undersized waveguide and for the lattice barrier, (T + R = 1) neglecting the small losses. The measured frequency band begins in the tunnelling frequency range (evanescent waves) and continues to the frequency regime above tunnelling (real waves). The curves display the undisturbed tunnelling barriers, where the reflection ($R_0$) is high, ≈ 1 (i.e. ≈ 0 dB return loss), while the transmission ($T_0$) is low in the tunnelling frequency range. The tunnelling regime ends near 10.2 GHz for the lattice (Fig.3) and near 9.5 GHz for the undersized waveguide (Fig.4), where the reflection drops with a minimum. On the other hand the transmission increases in the opposite way. The peaks are due to the singularity at the cut-off frequency [15]. Depending on the frequency, from below to above cut-off, the impedance is matched to that of the waveguide. $T_0$, $R_0$, $P_0$ were measured while the post was not inserted into the waveguide. By inserting the post, the negative and positive peaks of reflection and transmission are shifted to lower frequencies (i.e. reduced dielectric potential) at the evanescent to real wave transition. The post inside the waveguide thereby functions as a probe (only shown for 5 mm depth inside the waveguide).

The experimental data confirm the theoretical calculations: Obviously, evanescent modes and tunnelling, which are mathematically equivalent, can only be detected (become real) by supporting the wave packet with the missing barrier potential value or reducing the latter as expected [5].

The cavity model of Winful [27 – 29] explains the measured traversal time as a cavity decay time. Furthermore the input signal is not causally related with the output one. This strange property would exclude superluminal transmission of music [30] and more generally, for instance, the common application of tunnelling optical couplers in communication systems. These couplers are based on frustrated internal total reflection like the double prisms and described in its Minkowski diagram (see Fig. 1a and Fig. 5).

In conclusion, we measured the VSWR at microwave frequencies for two types of barriers, which yields a reflectivity of ≈ 0.98 within the experimental accuracy. The investigations have revealed: the quarter wavelength lattices have an inductive component and the undersized waveguides a capacitive component. The universal reflection time observed for symmetric barriers and the weak virtual transmission of opaque barriers did not influence the VSWR. (Opaque means that the product of the imaginary wave number κ and the barrier length z is ≥ 1.)

Experiments with reduced barrier dielectric potentials have shown that evanescent modes become transduced to detectable real waves. The new experimental results rule out the cavity as a tunnelling barrier. Especially, the claimed non-causal input to output relation was neither observed in experiments nor in fiber communication systems.

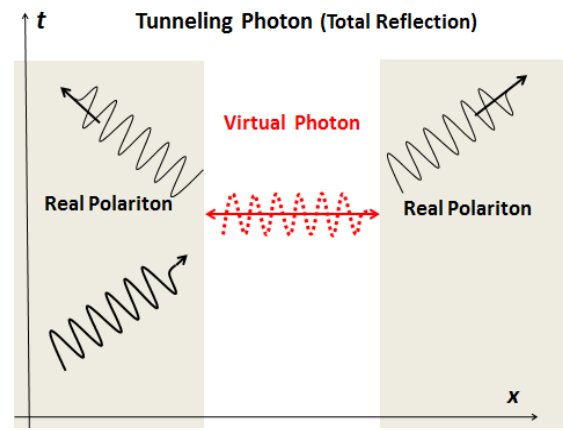

**Fig. 5:** Minkowski diagram (time t vs distance x) of a virtual photon in the barrier gap of double prisms under total internal frustrated reflection [4].

Remarkably, tunnelling processes and evanescent modes scale in the macro cosmos as well as at the quantum level (i.e. the tunnelling behaviour at quantum and macro scales is the same). Another interesting feature of the virtual photons, besides their non-local behaviour, is their space-like property as sketched in the Minkowski diagram of Fig. 5. In agreement with the Hartman effect, there is zero time spent inside an opaque barrier. The observed transmission and reflection times are equal and independent of barrier length in the case of symmetrical 1-dimensional barriers. Remarkably, the universal tunnelling time behaviour has also been observed for phonons [4]. Brillouin stated in his book on *Wave propagation in periodic structures* [31] that: the zone structure is completely independent of the special physical meaning of the waves considered, and it must be the same for elastic, electromagnetic, and Schrödinger electronic waves. Obviously, this statement also holds for the tunnelling process.

Many thanks to Bernhard Janssen for ongoing software support.